\documentclass[prd,twocolumn]{revtex4}

\usepackage{graphicx}
\usepackage{hyperref}
\usepackage{axodraw}

\renewcommand{\slash}{\not}

\makeatletter
\@addtoreset{equation}{section}

\makeatother

\newcommand{\beq}{\begin{equation}}
\newcommand{\eeq}{\end{equation}}

\newcommand{\di}{\displaystyle}

\newcommand{\ga}{\gamma}

\newcommand{\la}{\lambda}

\newcommand{\si}{\sigma}
 
\newcommand{\ve}{\varepsilon}

\newcommand{\PPP}{P_{33}(1232)}
\newcommand{\PP}{P_{11}(1440)}
\newcommand{\DD}{D_{13}(1520)}
\renewcommand{\SS}{S_{11}(1535)}

\newcommand{\pk}{{k \cdot p}}
\newcommand{\pkprime}{{k' \cdot p}}

\newcommand{\pq}{{q\cdot p}}
\newcommand{\W}{{p'}}

\newcommand{\Wq}{{q \cdot p'}}

\newcommand{\GeV}{\; {\mathrm{GeV}}}
\newcommand{\cm}{\; {\mathrm{cm}}}

\begin{document}

\title{
Resonance production by neutrinos: The second resonance region.
}
\author{Olga Lalakulich}
\email[E-mail:]{olalakul@zylon.physik.uni-dortmund.de}
\author{Emmanuel A. Paschos}
\email[E-mail:]{paschos@physik.uni-dortmund.de}
\author{Giorgi Piranishvili}
\affiliation{Institut of Physics, Dortmund University, 44221, Germany}
\date{\today}

\begin{abstract}
The article contains new results for spin-3/2 and -1/2 resonances. It specializes to the second resonance region, which includes the $\PP$, $\DD$ and $\SS$ resonances. New data on electroproduction enable us to determine the vector form factors accurately. Estimates for the axial couplings are obtained from decay rates of the resonances with the help of the partially conserved axial current (PCAC) hypothesis. We present cross sections to be compared with the running and future experiments. The article is self--contained and allows the reader to write simple programs for reproducing the cross sections and for obtaining additional differential cross sections.
\end{abstract}


\maketitle

\section{Introduction}

In previous articles \cite{Paschos:2003qr,Lalakulich:2005cs} we described the formalism for the excitation of the $P_{33}(1232)$ resonance. In the meanwhile, we extended the analysis to the second resonance region, which includes three isospin $1/2$ states:  $P_{11}(1440)$, $D_{13}(1520)$, $S_{11}(1535)$. The dominant $P_{33}(1232)$  has been observed in neutrino reactions and there are several theoretical articles which describe it with dynamical calculations based on unitarized amplitudes through dispersion relations \cite{Adler:1968tw}, phenomenological \cite{Schreiner:1973mj,Alvarez-Ruso:1998hi,Hagiwara:2003di,Amaro:2004bs} and quark models \cite{Rein:1980wg,Kuzmin:2003ji,Kuzmin:2004ya}, as well as models incorporating mesonic states \cite{Sato:2003rq} including a cloud of pions. Articles in the past five years have taken a closer look at the data by analysing the dependence on neutrino energy and $Q^2$ \cite{Alvarez-Ruso:1998hi,Paschos:2003qr,Lalakulich:2005cs}. So far a consistent picture emerged to be tested in the new accurate experiments. 

For the higher resonances there are several articles, that describe their excitation by electrons \cite{Ochi:1997ev,Armstrong:1998wg,Kamalov:2001yi,Yang:2005cq}, and only one \cite{Rein:1980wg} by neutrinos.  Experimental data for neutrino excitation of these resonances are very scarce and come from old bubble--chamber experiments \cite{Barish:1978pj,Radecky:1981fn,Kitagaki:1986ct,Grabosch:1988gw,Allasia:1990uy}. 
In the new experiments, studying neutrino oscillations, there is a strong
interest to go beyond the QE scattering \cite{Morfin:2005yt,Morfin:2005yp} and understand the excitation of these resonances. One reason comes from the long--baseline experiments where the two detectors (nearby and faraway) observe different regions of neutrino fluxes and kinematic regions of the produced particles.

A basic problem with resonances deals with the determination of their form factors (coupling constants and $Q^2$ dependences). The problem was apparent in the $\Delta$ resonance where  after many years several of the form factors and their $Q^2$ dependence became accurately known and were found to deviate from the dipoles. The situation is more serious for the higher resonances where the results of specific models are used. In this article we adopt the approach of determining the vector couplings from helicity amplitudes of electroproduction data, which became recently available from the Jefferson Laboratory \cite{Burkert:2002zz,Burkert:2004sk,Aznauryan:2004jd} and Mainz accelerator  \cite{Tiator:2003uu}. This requires that we write amplitudes for electroproduction in terms of the electromagnetic form factors and then relate them to the vector form factors that we use in neutrino reactions.  The above approach together with CVC uniquely specifies the couplings and the $Q^2$-dependences in the region of $Q^2$ where data is available, that is for $Q^2<3.5 \GeV^2$. 

The axial form factors are more difficult to determine. For the axial form factors we adopt an effective Lagrangian for the $R \rightarrow N \pi $ couplings and calculate the decay widths. For each resonance we assume PCAC which gives us one relation and a second coupling is determined using the decay width of each resonance. 

Having determined the couplings for the four resonances, we are able to calculate differential  and integrated cross sections. This way we investigate several properties in the excitation of the resonances. We find that a second resonance peak with an invariant mass between $1.4\GeV$ and $1.6 \GeV$ should be observable provided that neutrino energy is larger than $2-3\GeV$. Calculating cross sections in terms of the resonances provides a benchmark for their contribution and allows investigators to decide, when more precise data becomes available, whether a smooth background contribution is required. Integrated cross section already suggest the presence of a background.

In sections~\ref{electro-hel-amp}, \ref{isospin-relat} and \ref{resonances} we present the formalism and give expressions for the vector form factors. Estimates for the axial form factors are presented in the Appendix~\ref{gold-trei}. Section~\ref{tau} points out that the structure functions ${\cal W}_4$ and  ${\cal W}_5$ are important for reactions with tau neutrinos. We analyze differential and integrated cross sections in section~\ref{xsec-2nd}. We discuss there the existing data and point out a discrepancy in the normalization to be resolved in the next generation of experiments.


\section{Electroproduction via helicity amplitudes \label{electro-hel-amp}}

One of the main contributions of this article is the determination of the vector form factors for weak processes. Our work relates form factors to electromagnetic helicity amplitudes, whose numerical values are available from the Jefferson Laboratory and the University of Mainz. Values for the form factors at $Q^2=0$ are presented in Table~\ref{res-couplings}. Later on we express the weak structure functions in terms of form factors.

In applying this method we must still define the normalization of electromagnetic amplitudes, which is done in this section. Before we address this topic we discuss the kinematics, the polarizations and spinors entering the problem.

\begin{figure}[th]
\includegraphics[angle=-90,width=0.7\columnwidth]{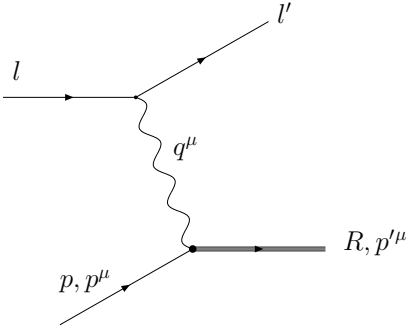}
\caption{Electroproduction of the resonance.}
\label{Fey-dia}
\end{figure}

We shall calculate Feynman amplitudes shown in Fig.\ref{Fey-dia}. We estimate the amplitudes in the laboratory frame with the initial nucleon at rest and with the intermediate photon moving along the $z-$axis. 

We define the four--momenta
\[
p^{\mu}=(m_N; 0,0,0), \qquad q^{\mu}=(q^0; 0,0,q^z) 
\]
\[
\mbox{and}\quad \W^{\mu}=p^\mu + q^\mu = (q^0+m_N; 0,0,q^z) .
\]
The intermediate photon or $W-$boson can have three polarizations $\varepsilon^{\mu}_{(i)}$ defined as
\beq
\begin{array}{c} \di
\varepsilon^\mu_{(R,L)}=\mp \frac1{\sqrt{2}}\bigl(0; 1, \pm i, 0), \qquad
\\[5mm] \di
\varepsilon^\mu_{(S)}=\frac1{\sqrt{Q^2}}\bigl(q^z; 0, 0, q^0) 
\end{array}
\eeq
and $Q^2=-q^2$.
The spinors for the target nucleon, normalized as $\bar u (p,s_z) u (p,s_z) = 2m_N$, are given by
\beq \di
u(p,s_z)=\sqrt{2m_N}\left(
\begin{array}{c} u_{s_z} \\  0 \end{array} \right)
\label{spinor-1/2}
\eeq
where $u_{s_z}$ can be either
\[ \di
u_{+1/2}= \left(  \begin{array}{c} 1 \\ 0 \end{array} \right) 
\quad \mbox{or} \quad 
u_{-1/2}= \left(  \begin{array}{c} 0 \\ 1 \end{array} \right). 
\]
The final resonances will have spin $1/2$ or $3/2$. For  spin-3/2 resonances we shall use Rarita--Schwinger spinors constructed as the product of a polarization vector $e^{\mu}_{(i)}$ with a spinor $U$. The states with various helicities are defined by
\beq
\begin{array}{l}
\psi_\mu^{(3/2)}=
e_\mu^{(R)} U(\W, +1/2)
\\[3mm]
\psi_\mu^{(1/2)}
=\sqrt{\frac23} e_\mu^{(S)} U(\W, +1/2) + \sqrt{\frac13} e_\mu^{(R)} U(\W, -1/2)
\\[3mm]
\psi_\mu^{(-1/2)}=\sqrt{\frac23} e_\mu^{(S)} U(\W, -1/2) + \sqrt{\frac13} e_\mu^{(L)} U(\W, +1/2)
\\[3mm]
\psi_\mu^{(-3/2)}=e_\mu^{(L)} U(\W, -1/2)
\end{array}
\label{spinvector-3/2}
\eeq 
with the spinor given as 
\[ \di
U(p',s_z)=\sqrt{\W^0+M_R}\left(
\begin{array}{c} u_{s_z} \\   \frac{ \vec{\W}\cdot \vec{\sigma} }{\W^0+M_R} u_{s_z} \end{array} \right),
\]
and the polarization vectors by
\[
e^\mu_{(R,L)}=\mp \frac1{\sqrt{2}}\bigl(0; 1, \pm i, 0), \qquad
e^\mu_{(S)}=\frac1{M_R}\bigl(p'{}^z; 0, 0, p'{}^0) . 
\]
We emphasize that  $\varepsilon^{\mu}_{(i)}$ refer to the intermediate photon and  $e^{\mu}_{(i)}$ belong to  the $J=3/2$ spin state.
For spin-1/2 resonances the spinors are 
\beq \di
u(p',s_z)=\sqrt{\W^0+M_R}\left(
\begin{array}{c} u_{s_z} \\   \frac{ \vec{\W}\cdot \vec{\sigma} }{\W^0+M_R} u_{s_z} \end{array} \right) .
\label{spinor-resonance}
\eeq

With this notation we can calculate three helicity amplitudes for the electromagnetic process. For instance, for the $D_{13}$ resonance the amplitude $\langle R,+1/2|J_{em}\cdot \ve_{(R)}| N,-1/2\rangle$ in terms of form factors is presented in eqs.~(\ref{vertexD13V}) and (\ref{D1520-A12}). 

We now define the overall normalization. Analyses of electroproduction data give numerical values for cross sections at the peak of each resonance \cite{Burkert:2002zz,Gorchtein:2004jd,Aznauryan:2004jd,Tiator:2003uu}
\begin{equation}
\sigma_{T}(W \! = \! M_R)=\frac{2m_N}{M_{R}\Gamma_{R}}(A^2_{1/2}+A^2_{3/2}),
\label{sigmaTviaA}
\end{equation}
\begin{equation}
\sigma_{L}(W \! = \! M_R)=\frac{2m_N}{M_{R}\Gamma_{R}}\frac{Q^2}{q^2_z}S^2_{1/2} .
\label{sigmaLviaA}
\end{equation}

These are helicity cross sections for the absorption of the "virtual" photon by the nucleon to produce the final resonance \cite{Hand:1963bb,Bjorken:1969ja}. They are defined as 
\begin{eqnarray}\nonumber
\sigma_{(i)}(W)=\frac12 K \sum\limits_{\lambda, \, s}\left| \langle R,\la| \varepsilon^\nu_{(i)} J^{el}_{\nu}| N,s \rangle \right|^2
R(W,M_R)
\label{sigmai}
\end{eqnarray}
\[
\mbox{with} \quad K=\frac{4\pi^{2}\alpha}{W^2-m_N^2} . 
\]
The last factor in the cross section is the Breit--Wigner term of a resonance:
\beq
R(W,M_R)=\frac{M_R \Gamma_R}{\pi} \frac{1}{(W^2-M^2_R)^2+M_R^2 \Gamma_R^2} .
\label{delta-function}
\eeq
For a very narrow resonance or a stable particle it reduces to a $\delta-$function.

Numerical values have been reported for the amplitudes in (\ref{sigmaTviaA}), (\ref{sigmaLviaA}) which are related to the following matrix elements
\beq
\begin{array}{l} \di
A_{1/2}  =  \sqrt{\frac{\pi\alpha}{m_N(W^2-m_N^2)}}\langle
R,+\frac{1}{2}|J_{em}\cdot \ve^{(R)}| N,-\frac{1}{2}\rangle ,
\\[4mm] \di
A_{3/2}  = \sqrt{\frac{\pi\alpha}{m_N(W^2-m_N^2)}}\langle
R,+\frac{3}{2}|J_{em}\cdot \ve^{(R)}| N,+\frac{1}{2}\rangle ,
\\[4mm] \di
S_{1/2} = \sqrt{\frac{\pi\alpha}{m_N(W^2-m_N^2)}} \frac{q_z}{\sqrt{Q^2}} \langle 
R,+\frac{1}{2}|J_{em}\cdot \ve^{(S)}| N,+\frac{1}{2}\rangle ,
\end{array}
\label{helic-elem}
\eeq
which we calculate in this article.

Following standard rules for the calculation of the expectation values the signs in these equations are determined. There is an ambiguity for the sign of the square root, which we select for all resonances to be positive. Later on we must also select the sign for axial form factors. We shall choose them in such a way that the structure functions $W_3$ for all resonances are positive, as indicated or suggested by the data. As a consequence the neutrino induced cross sections are larger than the corresponding antineutrino cross sections.

\section{Isospin  relations between electromagnetic and weak vertices \label{isospin-relat}}

Our aim is to relate the electromagnetic to weak form factors using isotopic symmetry. The photon has two isospin components  $|I,I_3\rangle=|1,0\rangle$ and $|0,0\rangle$. The isovector component belongs to the same isomultiplet with the vector part of the weak current. Each of the amplitudes $A_{3/2}$, $A_{1/2}$, $S_{1/2}$ can be further decomposed into three isospin amplitudes. Let us use a general notation and denote by $b$ the contribution from the isoscalar photon; similarly  $a^1$ and $a^3$ denote contributions of isovector photon to resonances with isospin $1/2$ and $3/2$, respectively.
A general helicity amplitude on a proton ($A_p$) and neutron ($A_n$) target has the decomposition
\begin{eqnarray} \di
A_{p}&=&A_{p}(\gamma p \to R^{+}) = b-\sqrt{\frac13}a^{1} + \sqrt{\frac23} a^3,  \nonumber
\\[4mm] \di
A_{n}&=&A_{n}(\gamma n \to R^{0}) = b+\sqrt{\frac13}a^{1} + \sqrt{\frac23} a^3 .
\label{p1}
\end{eqnarray}

For the weak current we have only an isovector component of the vector current, therefore the $b$ amplitude never occurs in weak interactions. A second peculiarity of the charged current is that $V_1\pm i V_2$ does not have the normalization for the Clebsch--Gordon coefficients, it must be normalized as $(V_1\pm i V_2)/\sqrt{2}$, which brings an additional factor of $\sqrt{2}$ to each of the charged current in comparison with the Clebsch--Gordon coefficients:
\beq
A(W^+ n \to R^{(1)+}) = \frac{2}{\sqrt{3}}a^1, 
\label{w1}
\eeq

\beq
\begin{array}{c} \di
A(W^+ p \to R^{(3)++}) = \sqrt{2}a^3, 
\\[4mm] \di
A(W^+ n \to R^{(3)+}) = \sqrt{\frac23}a^3 .
\end{array}
\label{w3}
\eeq

Comparing (\ref{p1}) with (\ref{w1}), one easily sees,
that, for the isospin-1/2 resonances, the weak amplitude satisfies the 
equality $A(W^+ n \to R^{(1)+}) = A_n^{1} - A_p^{1}$.
Since the amplitudes are linear functions of the form factors, the weak vector form factors are related in the same way to electromagnetic form factors for neutrons $C_i^n$ and protons $C_i^p$:
\beq 
I=1/2: \quad C_i^V=C_i^{n}-C_i^{p},
\label{iV-1/2}
\eeq 
with index $i$ distinguishing the Lorenz structure of the form factors. 

For the isospin-3/2 resonances one gets
$A_n^3(W^+ n \to R^{(3)+}) =  A_p^3(W^- p \to R^{(3)0})=\sqrt{2/3}a^3$.
The weak form factors, which are conventionally specified for these two processes, are 
\beq 
I=3/2: \quad C_i^V=C_i^{p}=C_i^{n}.
\label{iV-3/2}
\eeq 
For the process $W^+ p \to R^{(3)++}$ the amplitude is $\sqrt{3}$ times bigger: $A(W^+ p \to R^{(3)++})=\sqrt{3}A(W^+ n \to R^{(3)+})$. Some of the above relations were explicitly used in earlier articles \cite{Schreiner:1973mj,Paschos:2003qr,Lalakulich:2005cs}.

\section{Matrix elements of the resonance production, form factors \label{resonances}}

Following the notation of our earlier article \cite{Lalakulich:2005cs}, we write the cross section for resonance production in a notation close to that of DIS, that is we express it in terms of the  hadronic structure functions. In the present notation the sign in front of ${\cal W}_3$ in the cross section (\ref{cross-sec}) is plus, 
\begin{eqnarray} \di
\frac{d\si^{\nu N}}{dQ^2 dW}=\frac{G^2}{4\pi}\cos^2\theta_C\frac{W}{m_N E^2} \Biggl\{
{\cal W}_1(Q^2+m_\mu^2) \nonumber
\\  \di
+\frac{{\cal W}_2}{m_N^2}\left[ 2(\pk)(\pkprime) - \frac12 m_N^2 (Q^2+m_\mu^2)\right]
\nonumber 
\\ \di \nonumber 
+\frac{{\cal W}_3}{m_N^2} \left[ Q^2\pk - \frac12\pq (Q^2+m_\mu^2) \right]
\\ \di  
+\frac{{\cal W}_4}{m_N^2}m_\mu^2 \frac{(Q^2+m_\mu^2)}{2}
-2\frac{{\cal W}_5}{m_N^2} m_\mu^2 (\pk) \Biggr\} ,
\label{cross-sec}
\end{eqnarray}
which implies that cross section for a reaction with neutrino exceeds that with antineutrino if the structure function ${\cal W}_3$ is positive.
The corresponding formula for electroproduction is obtained by replacing the overall factor $G^2\cos^2\theta_C/4\pi$ by $2\pi \alpha^2/Q^4$ and using the electromagnetic structure functions ${\cal W}_1^{em}$ and ${\cal W}_2^{em}$ instead of weak ones. In this case the contribution from ${\cal W}_4^{em}$ and ${\cal W}_5^{em}$ is negligible and ${\cal W}_3^{em}=0$.

\begin{widetext}
\begin{table}[hb]
\arraycolsep=3mm
\doublerulesep=1mm
\caption{Vector and axial couplings for the excitation of resonances ($Q^2=0$)} 
\[
\begin{array}{ccccccccccc} 
\hline
R & M_R, \GeV  & \Gamma_{tot}, \GeV & \mbox{elast} & g_{\pi N R} & \mbox{} & C_5^A & \mbox{} & C_3^V & C_4^V & C_5^V
\\ \hline
P_{33}(1232) & 1.232 & 0.120 & 0.995 & 15.3 & \mbox{} & 1.2 & \mbox{} & 2.13 & -1.51 & 0.48 
\\[1mm]
\DD          & 1.520& 0.125  & 0.5   & 19.0 & \mbox{} & -2.1  & \mbox{} & -4.08  &  1.51 & 0.31 
\\[4mm] 
\hline
R & M_R, \GeV  & \Gamma_{tot}, \GeV & \mbox{elast} & g_{\pi N R} & \mbox{} & g_1^A & \mbox{} & g_1^V & g_2^V & 
\\ \hline
\PP & 1.440 & 0.350 & 0.6 & 10.9 & \mbox{} & -0.51 &  \mbox{}  & -4.6 & 1.52
\\[1mm]
\SS & 1.535 & 0.150 & 0.4 & 1.12 & \mbox{} & -0.21 &  \mbox{}  & -4.0 & -1.68
\\
\hline 
\end{array}
\]
\label{res-couplings}
\end{table}
\end{widetext}

In the following subsections we specify the structure functions ${\cal W}_i$ for each resonance by relating them to form factors.
To give the reader a quick overview of the results, we summarize the couplings (the  values of the form factor at $Q^2=0$) in Table~\ref{res-couplings}. One should keep in mind, however, that all form factors have different $Q^2$--dependences, which are given explicitly in  the following subsections.

\subsection{Resonance $D_{13}(1520)$}

We begin with a $D_{13}$ resonance, which has spin-3/2 and negative parity. The matrix element of the charged current for the resonance production is expressed as
\beq
\langle D_{13} | J^\nu | N \rangle = \bar \psi_\la^{(D)} (\W) d^{\la\nu}_{D_{13}} u(p)
\label{d13}
\eeq
with $u(p)$ the spinor of the target and $\psi_\la^{(D)}$ the Rarita-Schwinger field for a $D_{13}$ resonance. The structure of $d^{\la\nu}_D$ is given in term of the weak form factors

\begin{widetext}
\begin{eqnarray} \di 
d^{\la\nu}_{D_{13}}&=& \di g^{\la\nu}\left[ \frac{C_3^V}{m_N} \slash{q}  + \frac{C_4^V}{m_N^2}(p'q) 
                 + \frac{C_5^V}{m_N^2} (pq)  + C_6^V \right]
-q^\la \left[ \frac{C_3^V}{m_N} \ga^\nu  + \frac{C_4^V}{m_N^2} p'{}^\nu 
            + \frac{C_5^V}{m_N^2} p^\nu  \right]  \nonumber
\\[5mm] \hspace*{12mm} \di
&+& \di g^{\la\nu}\left[ \frac{C_3^A}{m_N} \slash{q} + \frac{C_4^A}{m_N^2} (p'q)  \right] \ga_5
- q^\la \left[ \frac{C_3^A}{m_N} \ga^\nu  + \frac{C_4^A}{m_N^2} p'{}^\nu  \right] \ga_5
+ \left[ g^{\la\nu} C_5^A  + q^\la q^\nu \frac{C_6^A}{m_N^2} \right] \ga_5. 
\label{dlanu-d13}
\end{eqnarray}
\end{widetext}

The general form of the current for $D_{13}$ differs from that of $P_{33}$ in the location of the  $\ga_5$ matrix, which is due to the parity of the resonance. The form factors $C^V_i(Q^2)$ and $C^A_i(Q^2)$ refer now to any  $D_{13}$ resonance. 
Later we'll specify them for the $\DD$.  The vector form factors are extracted from the electroproduction data, in particular from the helicity amplitudes. We use recent data from \cite{Aznauryan:2004jd,Aznauryan:2005oral}, which were kindly provided to us by I. Aznauryan.

Helicity amplitudes are expressed via the $J_{em}\cdot \varepsilon$, which  can be obtained from (\ref{d13}) and (\ref{dlanu-d13}) by setting the axial couplings equal to zero and replacing the vector form factors by the electromagnetic form factors. This results in the following expression 

\begin{eqnarray}
\di
\langle D_{13}| J_{em}\cdot \varepsilon |N \rangle &=&
\bar{\psi}_{\mu}  \Gamma_\nu^{(em)} \, F^{\mu\nu} u(N),
\nonumber \\[2mm] 
\mbox{with} \qquad \Gamma_\nu^{(em)}&=&\frac{C_3^{(em)}}{m_N}\gamma_{\nu}
        +\frac{C_4^{(em)}}{m_N^2} p^{\prime}_{\nu}+\frac{C_5^{(em)}}{m_N^2}p_{\nu}
\nonumber \\[2mm] 
F^{\mu\nu}&=& q^\mu\varepsilon^\nu - q^\nu\varepsilon^\mu  
\label{vertexD13V}
\end{eqnarray}

As it was discussed in the previous section, the electromagnetic form factors of $D_{13}$ resonance are different for proton and neutron. Substituting the matrix element (\ref{vertexD13V}) into Eqs. (\ref{helic-elem}) and carrying out the products with the spinors and Rarita--Schwinger field we obtain the following helicity amplitudes for electroproduction

\begin{eqnarray} \di
A_{3/2}^{D_{13}}&=& \sqrt{N}
\biggl[
\frac{C_3^{(em)}}{m_N}(M_R-m_N)
\nonumber \\[1mm] \di
&+&\frac{C_4^{(em)}}{m_N^2}\Wq
+\frac{C_5^{(em)}}{m_N^2}\pq
\biggr]
\label{D1520-A32}
\end{eqnarray}

\begin{eqnarray} \di
A_{1/2}^{D_{13}}&=&\sqrt{\frac{N}{3}}
\biggl[
\frac{C_3^{(em)}}{m_N}(M_R-m_N -\frac{2m_N}{M_R}\frac{q_z^2}{\W^0+M_R}) 
\nonumber \\[1mm] \di
&+&\frac{C_4^{(em)}}{m_N^2}\Wq
+\frac{C_5^{(em)}}{m_N^2}\pq 
\biggr] \phantom{ccccccc}
\label{D1520-A12}
\end{eqnarray}

\begin{eqnarray} \di
S_{1/2}^{D_{13}}&=&\sqrt{\frac{2N}{3}}\frac{q^z}{M_R} 
\biggl[
\frac{C_3^{(em)}}{m_N}\left( -M_R \right) +\frac{C_4^{(em)}}{m_N^2}(Q^2
\nonumber  \\[1mm] \di  
&-&2m_N q^0-m_N^2)
-\frac{C_5^{(em)}}{m_N}(q^0+m_N)
\biggr]  
\label{D1520-S12}
\end{eqnarray}
where $\di  N= \frac{\pi\alpha_{em}}{m_N(W^2-m_N^2)}2m_N(p'{}^0+M_R)$.

\begin{figure}[ht]
\begin{center}
\includegraphics[angle=-90,width=\columnwidth]{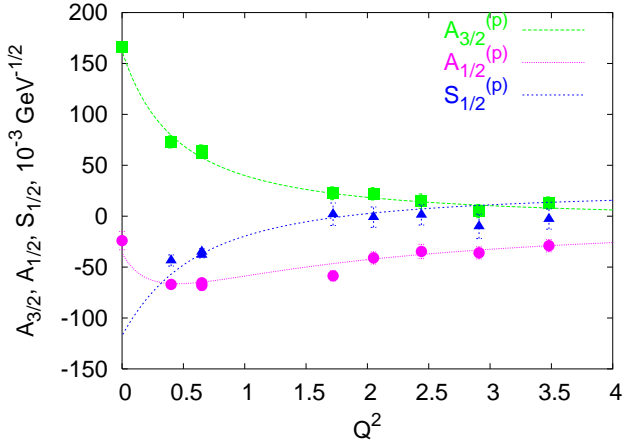} 
\end{center}
\caption{Helicity amplitudes for the $\DD$ resonance, calculated with the form factors from Eq.(\ref{ff-D1520}).}
\label{fig-D1520-AS}
\end{figure}

Comparing Eqs. (\ref{D1520-A32}), (\ref{D1520-A12}), (\ref{D1520-S12}) for each value of $Q^2$ with the recent data on helicity amplitudes \cite{Burkert:2002zz,Aznauryan:2004jd,Aznauryan:2005oral}, we extract the following form factors:

\begin{equation}
\begin{array}{cl} \di 
D_{13}(1520): \quad &  \di C_3^{(p)}=\frac{2.95/D_V}{1+Q^2/8.9 M_V^2},
\\ &	\di 
C_4^{(p)}=\frac{-1.05/D_V}{1+Q^2/8.9 M_V^2},
	\\ & \di	
C_5^{(p)}=\frac{-0.48}{D_V} .
\\[4mm] 
& \di C_3^{(n)}=\frac{-1.13/D_V}{1+Q^2/8.9 M_V^2},
	\\ & \di	
C_4^{(n)}=\frac{0.46/D_V}{1+Q^2/8.9 M_V^2}
	\\ & \di	
C_5^{(n)}=\frac{-0.17}{D_V} ,
\end{array}
\label{ff-D1520}
\end{equation}
This is a simple algebraic solution with the numerical values for the form factors being unique.
The function $D_V=(1+Q^2/M_V^2)^2$ denotes the dipole function with the vector mass parameter $M_V=0.84\GeV$. To give an impression, how good this parametrisation is, we plot in Fig.\ref{fig-D1520-AS}
the helicity amplitudes, obtained with these form factors.  Vector form factors for the neutrino--nucleon interactions are calculated according to Eq. (\ref{iV-1/2})

\begin{figure}[ht]
\includegraphics[angle=-90,width=\columnwidth]{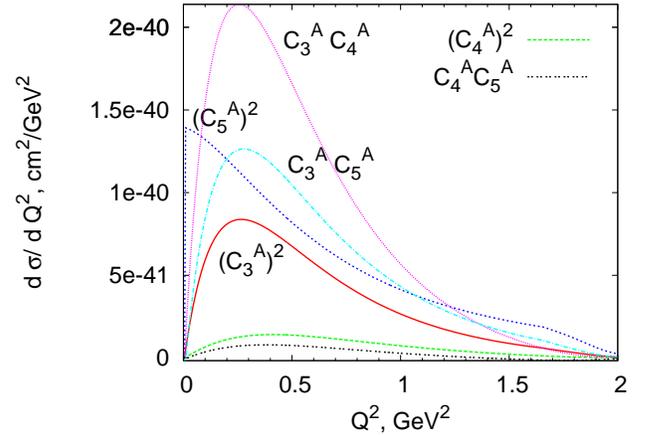}
\caption{Contribution of various form factors to the differential cross section.} 
\label{fig-difrFF}
\end{figure}

For the axial form factors we derive in the Appendix~\ref{gold-trei}
\beq
C_6^A{}^{(D)}=m_N^2 \frac{C_5^A{}^{(D)}}{m_\pi^2+Q^2},
\quad
C_5^A{}^{(D)}(0)=-2.1
\label{c6a-D13}
\eeq
Two other form factors and the $Q^2$ behaviour of the $C_5^A$ can be determined either experimentally or in a specific theoretical model. 
To check how big the contribution of the  $C_3^A$ and $C_4^A$ could be, we set them  $C_3^A=C_4^A=1$ and computed in Fig.\ref{fig-difrFF} the various contributions to the differential cross section for $E_\nu=2\GeV$. Motivated by the results on $\PPP$ resonance \cite{Paschos:2003qr}, the $Q^2$ dependence in our calculations is taken as  
\beq
C_i^A{}^{(D)} = \frac{C_i^A{}^{(D)}(0)/D_A}{1+Q^2/3M_A^2},
\eeq
where $D_A=(1+Q^2/M_A^2)^2$ denotes the dipole function with the axial mass parameter $M_A=1.05 \GeV$.

We conclude from Figure \ref{fig-difrFF}, that the contribution of $(C_4^A)^2$  and $C_4^AC_5^A$ are small, but the other terms could be sizable. Their importance depends on the relative signs.
It is possible, that $C_3^A C_4^A$ and $C_3^A C_5^A$ are positive and together with $(C_3^A)^2$ give an additional $100\%$ to the $D_{13}$ cross section. In case that $C_3^A C_4^A$ and/or $C_3^A C_5^A$ are negative, there are cancellations.  In the following calculations of this article we set  for simplicity $C_3^A=C_4^A=0$. 

The hadronic structure functions for $D_{13}$ resonance are similar to those for $P_{33}$, presented in our earlier paper \cite{Lalakulich:2005cs} and can be obtained from them by replacing $m_N M_R$ by $-m_NM_R$. We repeat the corresponding formulas here including the terms with $C_5^V$ and $C_3^A$, which could be nonzero. The structure functions have a form 
\beq
{\cal W}_i(Q^2,\nu)=\frac{2}{3m_N} V_i(Q^2,\nu) R(W,M_R)
\eeq
where $R(W,M_R)$ was defined in (\ref{delta-function}) and $V_i$ are given below.
In the following equations the upper sign corresponds to $P_{33}$ resonance and the lower sign to the $D_{13}$. 

\begin{widetext}
\begin{eqnarray} \di  \nonumber 
V_1
&=& \frac{(C_3^V)^2}{m_N^2 M_R^2} 
          \left[(Q^2-\pq)^2(\pq+m_N^2)
	       + M_R^2( (\pq)^2 +Q^2 m_N^2 \pm Q^2 m_N M_R)
	  \right]
\\[3mm]  \di \nonumber
&+& \frac{(C_3^A)^2}{m_N^2 M_R^2} 
          \left[(Q^2-\pq)^2(\pq+m_N^2)
	       + M_R^2((\pq)^2+Q^2 m_N^2 \mp Q^2 m_N M_R)
	  \right]
\\[3mm]  \di \nonumber
&+& \frac{C_3^V C_4^V (Q^2-\pq) - C_3^V C_5^V  \pq}{m_N^3 M_R} 
\left[ (Q^2-\pq)(\pq +m_N^2 \mp 2m_N M_R)- M_R^2 \pq \right]
\\[3mm]   \di 
&+& \frac{C_3^A C_4^A (Q^2-\pq) -  C_3^A C_5^A m_N^2}{m_N^3 M_R} 
            \left[ (Q^2-\pq) \left(\pq + (M_R\pm m_N)^2 \right) - M_R^2 Q^2 \right]
\\[3mm]  \di 
&+& 
\frac{\left[ C_4^V (Q^2-\pq) - C_5^V \pq \right]^2 }{m_N^4} 
        (\pq+m_N^2 \mp m_N M_R)
+\left[  C_5^A -  \frac{ C_4^A (Q^2-\pq)}{m_N^2}  \right]^2  
\left[ \pq +m_N^2 \pm m_N M_R \right]
\nonumber
\label{calW1}
\end{eqnarray}


\begin{eqnarray} \di
V_2 &=& 
\frac{(C_3^V)^2 + (C_3^A)^2}{M_R^2} Q^2 \left[ \pq +m_N^2 +M_R^2 \right]  
+\frac{C_3^V C_4^V}{m_N M_R}  Q^2 \left[ \pq + (M_R \mp m_N)^2 \right]  
+\frac{C_3^A C_4^A}{m_N M_R}  Q^2 \left[ \pq + (M_R \pm m_N)^2 \right]  
\nonumber
\\[3mm] \di
&+&\frac{C_3^V  C_5^V}{m_N M_R } Q^2 \left[ \pq +(M_R\mp m_N)^2 +Q^2\right]
+\left( \frac{(C_4^V)^2}{m_N^2} 
          + \frac{(C_5^V)^2(Q^2+M_R^2)}{m_N^2 M_R^2} 
          +\frac{2 C_4^V C_5^V}{m_N^2}                   \right)Q^2 
   \left[ \pq +m_N^2 \mp m_N M_R \right]  
\nonumber
\\[3mm] \di
&+& C_3^A C_5^A \frac{m_N}{M_R} Q^2
+\left[ ({C_5^A})^2 \frac{m_N^2}{M_R^2} 
+ \frac{(C_4^A)^2}{m_N^2} Q^2 \right]
\left[  \pq+m_N^2 \pm m_N M_R \right] 
\label{calW2}
\end{eqnarray}

\begin{eqnarray} \di
V_3 &=& \di
2\frac{C_3^V C_3^A}{M_R^2} \left[ 2(Q^2-\pq)^2 +M_R^2(3Q^2-4\pq) \right]
+2\left[ \frac{C_4^V C_4^A}{m_N^2}(Q^2 -\pq) -C_4^V C_5^A \right] (Q^2-\pq)
\nonumber 
\\[3mm] \di
&+& 2\frac{C_5^V C_3^A \pq - C_4^V C_3^A(Q^2-\pq)}{M_R m_N} 
\left[2M_R^2 \mp 2 m_N M_R +Q^2 -\pq \right] 
+ 2\left[ C_5^V C_5^A -\frac{C_5^V C_4^A}{m_N^2}(Q^2-\pq) \right] \pq
\nonumber 
\\[3mm] \di
& + & 2 \left[C_3^V C_5^A \frac{m_N}{M_R} - 
   \frac{C_3^V\, C_4^A}{M_R m_N} (Q^2-\pq)
  \right]
\left(2 M_R^2 \pm 2 m_N M_R + Q^2 - \pq \right)
\label{calW3}
\end{eqnarray}

These are the important structure functions for most of the kinematic region. There are two additional structure functions, whose contribution to the cross section is proportional to the square of the muon mass.

\begin{eqnarray} \di \nonumber
V_4 &=& 
\frac{(C_3^V)^2}{M_R^2}  \left[ (2\pq-Q^2)(\pq+m_N^2) -M_R^2(m_N^2 \pm m_N M_R) \right]  
\\[3mm]  \nonumber
\di
&+& 
\frac{(C_3^A)^2}{M_R^2}  \left[ (2\pq-Q^2)(\pq+m_N^2) -M_R^2(m_N^2 \mp m_N M_R) \right]  
\\[3mm]  \nonumber
\di
&+&\left[   \frac{(C_4^V)^2 (2\pq-Q^2) }{m_N^2} 
          + \frac{(C_5^V)^2 (\pq)^2 }{m_N^2 M_R^2} 
          + 2\frac{C_4^V C_5^V}{m_N^2} \pq
   \right]
\left[ \pq +m_N^2 \mp m_N M_R \right]  
\\[3mm] \nonumber  
\di
&+& \frac{C_3^V C_4^V}{m_N M_R} \left[ (2\pq-Q^2)(\pq +m_N^2 \mp 2 m_N M_R)+ \pq M_R^2 \right]  
+\frac{C_3^V C_5^V}{m_N M_R} pq \bigl[2\pq +m_N^2 \mp 2 m_N M_R +M_R^2 +Q^2\bigr]  
\\[3mm]  \nonumber
\di
&+&
\left[ 
(C_5^A)^2 \frac{m_N^2}{M_R^2}  
+ \frac{(C_4^A)^2}{m_N^2}(2\pq-Q^2) 
+\frac{(C_6^A)^2}{m_N^2 M_R^2}\biggl((Q^2-\pq)^2+Q^2M_R^2\biggr) 
\right.
\\[3mm]  \di \nonumber
& & \hspace*{25mm} \left. +2 C_4^A C_5^A -2\frac{C_4^A C_6^A}{m_N^2}\pq -2\frac{C_5^A C_6^A}{M_R^2}(M_R^2+Q^2-\pq)
\right] 
\left[\pq +m_N^2 \pm m_N M_R \right]
\\[3mm]  \di \nonumber
&+& \frac{C_3^A C_4^A}{m_N M_R} \left[ (2\pq - Q^2)(\pq+m_N^2 \pm 2m_N M_R) + M_R^2 \pq \right]
\\[3mm]  \di 
&+& C_3^A C_5^A\frac{m_N}{ M_R} (2\pq+m_N^2   \pm 2m_N M_R)
-\frac{C_3^A C_6^A}{m_N M_R} \pq \biggl(Q^2 +(M_R\pm m_N)^2 \biggr)
\end{eqnarray}


\begin{eqnarray} \nonumber
V_5 &=&  
 \frac{(C_3^V)^2 + (C_3^A)^2}{M_R^2}\pq  \left[ \pq +m_N^2+ M_R^2 \right]               
+\frac{C_3^V C_5^V}{m_N M_R} \pq \left[ \pq + (M_R \mp m_N)^2 +Q^2 \right]
\\[3mm]  
\di  \nonumber
&+&\left[     \frac{(C_4^V)^2}{m_N^2} 
          + \frac{(C_5^V)^2(Q^2+M_R^2)}{m_N^2 M_R^2} 
          + 2\frac{C_4^V C_5^V}{m_N^2}
\right] 
\pq \left[ \pq +m_N^2 \mp m_N M_R \right]
+\frac{C_3^V C_4^V}{m_N M_R} \pq \left[ \pq + (M_R \mp m_N)^2      \right]
\\[3mm]  
\di  \nonumber
&+&\frac{C_3^A C_4^A}{m_N M_R} \pq \left[ \pq+ (M_R\pm m_N)^2 \right]
+C_3^A C_5^A\frac{m_N}{2M_R} \left[ 2\pq+ Q^2+ (M_R\pm m_N)^2 \right]
+\frac{C_3^A C_6^A}{2 m_N M_R} Q^2 \left[ Q^2+ (M_R\pm m_N)^2 \right]
\\[3mm] \di
&+&\left[ \frac{(C_4^A)^2}{m_N^2}\pq + (C_5^A)^2 \frac{m_N^2}{M_R^2}
                +C_4^A C_5^A  - \frac{C_4^A C_6^A}{m_N^2}Q^2 
               + \frac{C_5^A C_6^A}{M_R^2}(\pq - Q^2)
        \right] \left[ \pq +m_N^2 \pm m_N M_R \right] 
\label{calW5}
\end{eqnarray}
\end{widetext}

\subsection{Resonance $P_{33}(1232)$}

The method of extracting the form factors from the helicity amplitudes, described in the previous section is applicable to any resonance. The helicity amplitudes for the $\PPP$ resonance were calculated in a similar manner and we obtain the following results:
\begin{eqnarray} \di
A_{3/2}^{P_{33}}&=&-\sqrt{N}\frac{q^z}{\W^0+M_R}\left[
\frac{C_3^{(em)}}{m_N}(m_N+M_R)
\right. \nonumber \\[4mm] \di 
&+&\left. \frac{C_4^{(em)}}{m_N^2}\Wq
+\frac{C_5^{(em)}}{m_N^2}\pq
\right] 
\label{P1232-A32}
\end{eqnarray}

\begin{eqnarray} \di
A_{1/2}^{P_{33}}&=&\sqrt{\frac{N}{3}}
\biggl[
\frac{C_3^{(em)}}{m_N}\bigl(m_N+M_R 
-2\frac{m_N}{M_R}(\W^0+M_R)\bigr)
\nonumber \\[4mm] \di
&+&\frac{C_4^{(em)}}{m_N^2}\Wq
+\frac{C_5^{(em)}}{m_N^2}\pq 
\biggr] \frac{q^z}{\W^0+M_R} 
\label{P1232-A12}
\end{eqnarray}

\begin{eqnarray} \di
S_{1/2}^{P_{33}}&=&\sqrt{\frac{2N}{3}}\frac{q_z^2}{M_R(\W^0+M_R)} 
\biggl[
\frac{C_3^{(em)}}{m_N}M_R
\nonumber \\[4mm] \di
&+&\frac{C_4^{(em)}}{m_N^2} W^2
+\frac{C_5^{(em)}}{m_N^2}m_N(m_N+q^0) 
\biggr] \phantom{cccc}
\label{P1232-S12}
\end{eqnarray}

\begin{figure}[ht]
\begin{center}
\includegraphics[angle=-90,width=\columnwidth]{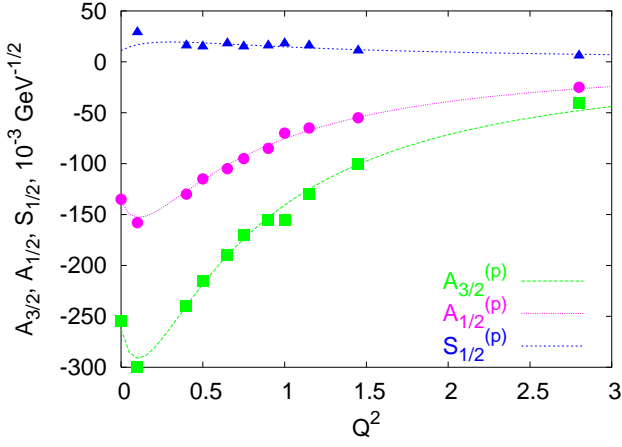} 
\end{center}
\caption{Helicity amplitudes for the $\PPP$ resonance, calculated with the form factors from Eq.(\ref{ff-P1232}). Data are from \cite{Tiator:2003uu}.}
\label{fig-P1232-AS}
\end{figure}

Comparing helicity amplitudes from Eqs. (\ref{P1232-A32}),(\ref{P1232-A12}),(\ref{P1232-S12}) with the available data \cite{Tiator:2003uu} allows us to extract the form factors
\begin{eqnarray} 
C_3^{(p)}=\frac{2.13/D_V}{1+Q^2/4 M_V^2},
\nonumber \\
C_4^{(p)}=\frac{-1.51/D_V}{1+Q^2/4 M_V^2}, 
\label{ff-P1232}
\\
C_5^{(p)}=\frac{0.48/D_V}{1+Q^2/0.776 M_V^2} .
\nonumber
\end{eqnarray}
Form factors $C_3^{(p)}$ and $C_4^{(p)}$ are in agreement with those obtained in the  magnetic dominance approximation (which was used in all the previous papers on neutrinoproduction). The agreement has $5\%$ accuracy and at the same time the nonzero  scalar helicity amplitude is described correctly. 
The fit of the proton helicity amplitudes for the form factors from Eq.(\ref{ff-P1232}) is shown in Fig.\ref{fig-P1232-AS}. 
Electromagnetic neutron form factors and vector form factors for the neutrino--nucleon interactions can be calculated according to Eq. (\ref{iV-3/2}).

Axial form factors have already been discussed several times, the way to obtain them is illustrated in Appendix~\ref{gold-trei}, Eq. (\ref{P1232-C5A}). The result is
\[
C_6^A{}^{\Delta}=m_N^2 \frac{C_5^A{}^{\Delta}}{m_\pi^2+Q^2}, 
\quad 
C_5^A{}^{\Delta}=\frac{1.2/D_A}{1+Q^2/3M_A^2}
\]


A practical aspect with this resonance concerns the cross section of the tau neutrino interactions, which is discussed in Section \ref{tau}.

\subsection{Resonance $P_{11}(1440)$}

For spin-1/2 resonances the parametrization  of the weak vertex for the resonance production is simpler than for the spin-3/2 resonances and is similar to the parametrization for quasi--elastic scattering.

The matrix elements of the $P_{11}$ resonance production can be written as follows:
\begin{eqnarray} \di
\langle P_{11} | J^\nu | N\rangle 
=\bar{u}(p')  \left[
\frac{g_1^V}{\mu^2}(Q^2\gamma^\nu + \slash{q} q^\nu) 
\right. 
\nonumber   
\\ \left. \di
+ \frac{g_2^V}{\mu} i \si^{\nu\rho} q_\rho
- g_1^A \ga^\nu \gamma_5 - \frac{g_3^A}{m_N} q^\nu \gamma_5\right] u(p), 
\label{meP1440}
\end{eqnarray}
where we use the standard notation $\si^{\nu\rho} = \frac{i}{2}[\ga^\nu, \ga^\rho]$ and the kinematic factors are scaled by $\mu=m_N+M_R$ in order to make them dimensionless.

To extract the vector form factors we use the same procedure as before and calculate the helicity amplitudes for the virtual photoproduction process. Since the resonance has spin $1/2$, only the $A_{1/2}$ and $S_{1/2}$ amplitudes occur:
\begin{eqnarray} \di
A^{P_{11}}_{1/2}=\sqrt{N} \frac{\sqrt{2} q^z}{p'{}^0+M_R}
\left[ \frac{g_1^{(em)}}{\mu^2}Q^2
\right.  \nonumber \\ \di \left.
+\frac{g_2^{(em)}}{\mu}(M_R+m_N)\right]
\label{P1440-A12}
\end{eqnarray}

\begin{eqnarray} \di
S^{P_{11}}_{1/2}=\sqrt{N} \frac{q_z^2}{p'{}^0+M_R} 
\left[\frac{g_1^{(em)}}{\mu^2}(M_R+m_N)
\right.  \nonumber \\ \di \left.
-\frac{g_{2}^{(em)}}{\mu} \right] 
\label{P1440-S12}
\end{eqnarray}

\begin{figure}[b!ht]
\begin{center}
\includegraphics[angle=-90,width=\columnwidth]{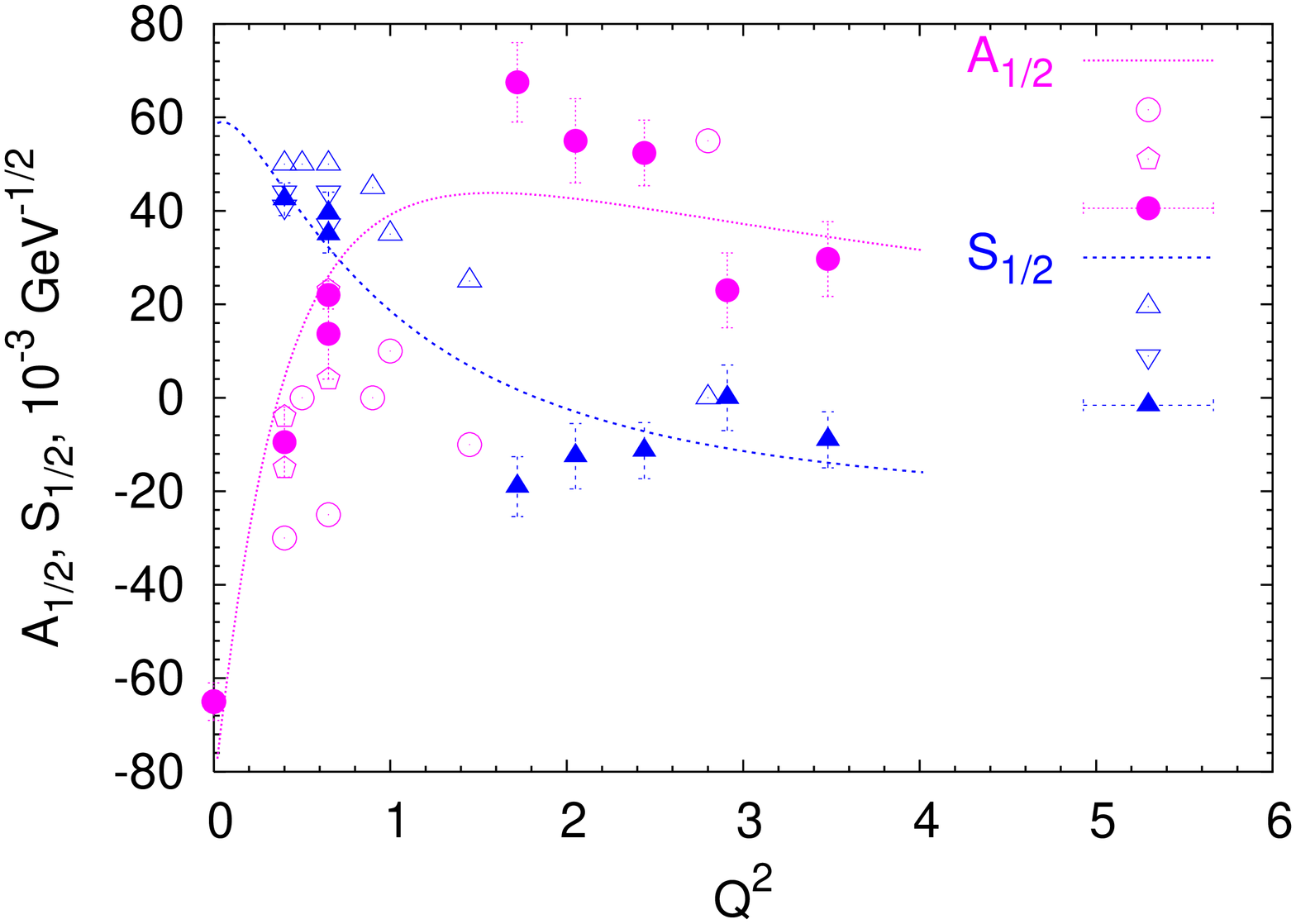} 
\end{center}
\caption{Helicity amplitudes for the $\PP$ resonance, calculated with the form factors from Eq.(\ref{ff-P1440}). For $A_{1/2}$ the ata are from:  \cite{Tiator:2003uu} (unshaded circles), \cite{Aznauryan:2004jd} (unshaded pentagons), \cite{Aznauryan:2005oral} (full circles);
for $S_{1/2}$: \cite{Tiator:2003uu} (unshaded up triangles), \cite{Aznauryan:2004jd} (unshaded down triangles), \cite{Aznauryan:2005oral} (full triangles)}
\label{fig-P1440-AS}
\end{figure}

At nonzero $Q^2$ data on helicity amplitudes for the $\PP$ are available only for proton. Unlike the other resonances, the accuracy of data in this case is low and numerical values, provided by different groups differ significantly, as  illustrated in Fig.\ref{fig-P1440-AS}. In this situation we use for our fit only the recent data \cite{Aznauryan:2004jd, Aznauryan:2005oral}. 

Matching the data against Eqs. (\ref{P1440-A12}), (\ref{P1440-S12}) allows us to parametrize the proton electromagnetic form factors as follows:
\begin{equation}
\begin{array}{ll}
\di P_{11}(1440): & \di
g_1^{(p)} = \frac{2.3/D_V}{1+Q^2/4.3 M_V^2}, 
\\ & \di
g_2^{(p)} = \frac{-0.76}{D_V}
	     \left[1 - 2.8 \ln\left(1+\frac{Q^2}{1\GeV^2}\right)
	     \right]\ .
\end{array}
\label{ff-P1440}
\end{equation}

The difference among the reported helicity amplitudes are larger than the estimated contribution of the isoscalar part of the electromagnetic current. For this reason we shall assume that the isoscalar contribution is negligibly small and use the relation $A^{(n)}_{1/2}=-A^{(p)}_{1/2}$. The isovector contribution in the neutrinoproduction is now given as $g_i^V=-2g_i^{(p)}$.

The differential cross section is expressed again with the general formula (\ref{cross-sec}). The hadronic structure functions are calculated explicitly to be:
\[
{\cal W}_i(Q^2,\nu)=\frac{1}{m_N} V_i(Q^2,\nu) R(W,M_R)
\]
\begin{eqnarray} \di
V_1&=&
\frac{(g_1^V)^2}{\mu^4} Q^4 \left[ (pq+m_N^2\mp m_N M_R) \right]
\nonumber \\ \di
&+&\frac{(g_2^V)^2}{\mu^2} \left[ 2(pq)^2+Q^2(m_N^2\pm m_N M_R-\pq)  \right]
\nonumber \\ \di
&+&\frac{g_1^V g_2^V}{\mu^3} 2 Q^2 \left[ (pq)(M_R\mp m_N) \pm m_N Q^2  \right]
\nonumber \\ \di
&+& (g_1^A)^2(m_N^2\pm m_N M_R+\pq) \phantom{cccccc}
\end{eqnarray}

\beq
V_2=2 m_N^2 \left[\frac{(g_1^V)^2}{\mu^4}Q^4 
+ \frac{(g_2^V)^2}{\mu^2}Q^2 + (g_1^A)^2 \right]
\eeq

\beq
V_3= 4 m_N^2 \left[ 
\frac{g_1^V g_1^A}{\mu^2}Q^2
+\frac{g_2^V g_1^A}{\mu}(M_R\pm m_N)
\right]
\eeq

\begin{eqnarray} \di
V_4 & = &
m_N^2\biggl[ 
\frac{(g_2^V)^2}{\mu^2} \left[ \pq -m_N^2 \mp m_N M_R \right]
\nonumber \\ \di
&+&\frac{(g_1^V)^2}{\mu^4} \left[ 2(pq)^2 - Q^2(pq+m_N^2\mp m_N M_R) \right]
\nonumber \\ \di
&-&\frac{g_1^V g_2^V}{\mu^3} \left[ \pq(M_R \mp m_N) \pm m_N Q^2 \right]
 \\ \di
&\mp& 2 g_1^A g_3^A + \frac{(g_3^A)^2}{m_N^2} \left[ (pq) + m_N^2 \mp m_N M_R  \right]  \biggr]
\nonumber
\end{eqnarray}

\begin{eqnarray} \di
V_5&=& m_N^2  \biggl[ 2\frac{(g_1^V)^2}{\mu^4}Q^2 \pq  
+2\frac{(g_2^V)^2}{\mu^2}\pq + 
\nonumber \\ \di
&+&(g_1^A)^2 + \frac{g_1^A g_3^A}{m_N}(M_R \mp m_N) \biggr] 
\end{eqnarray}
where the upper sign corresponds to $P_{11}$ and the lower sign to $S_{11}$ resonance.

As it is shown in Appendix~\ref{gold-trei}, PCAC allows us to relate the two axial form factors and fix their values at $Q^2=0$:
\[
g_3^A{}^{(P)}=\frac{(M_R+m_N)m_N}{Q^2+m_\pi^2} g_1^A{}^{(P)}, \quad  g_1^A{}^{(P)}(0)=-0.51 .
\]
The $Q^2$ dependence of the form factors cannot be determined by general theoretical consideration and will have to be extracted from the experimental data. We again suppose, that the form factors are modified dipoles  
\beq
g_1^A{}^{(P)}= \frac{g_1^A{}^{(P)}(0)/D_A}{1+Q^2/3M_A^2} .
\eeq

\subsection{Resonance $S_{11}(1535)$}

For the $S_{11}$ the amplitude of resonance production is similar to that for $P_{11}$ with the $\ga_5$ matrix exchanged between the vector and the axial parts
\begin{eqnarray} \di
\langle S_{11} | J^\nu | N\rangle 
=\bar{u}(p')  \left[
\frac{g_1^V}{\mu^2}(Q^2\gamma^\nu + \slash{q} q^\nu) \ga_5
\right. 
\nonumber   
\\ \left. \di
+ \frac{g_2^V}{\mu} i \si^{\nu\rho} q_\rho \ga_5
- g_1^A \ga^\nu  - \frac{g_3^A}{m_N} q^\nu \right] u(p) . 
\phantom{cccccc}
\label{meS1535}
\end{eqnarray}

The helicity amplitudes
\begin{eqnarray}
A^{S_{11}}_{1/2}=\sqrt{2N}
\left[
\frac{g_1^{(em)}}{\mu^2}Q^2 
+
\frac{g_{2}^{(em)}}{\mu}\left(M_R-m_N\right)\right]
\label{S1535-A12}
\end{eqnarray}

\begin{eqnarray} \nonumber
S^{S_{11}}_{1/2}=\sqrt{N}  q_z
\left[ -\frac{g_1^{(em)}}{\mu^2} \left(M_R-m_N\right) 
+\frac{g_{2}^{(em)}}{\mu}\right] 
\label{S1535-S12}
\end{eqnarray}
are used to extract the electromagnetic form factors.

As in the case of $\PP$ resonance, we choose here to fit only proton data on helicity amplitudes and neglect the isoscalar contribution to the electromagnetic current 
\begin{figure}[ht]
\begin{center}
\includegraphics[angle=-90,width=\columnwidth]{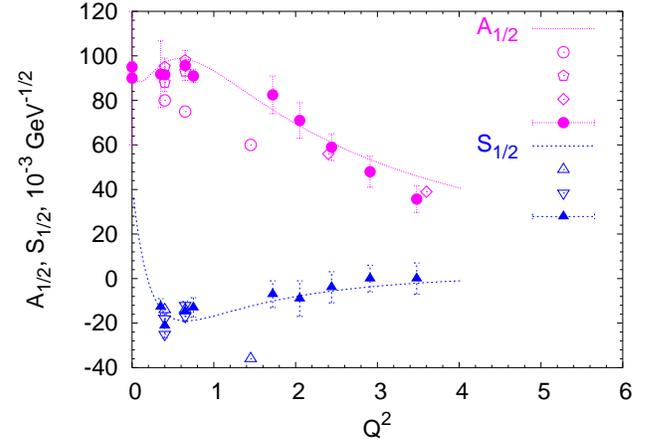} 
\end{center}
\caption{Helicity amplitudes for the $\PP$ resonance, calculated with the form factors from Eq.(\ref{ff-S1535}). For $A_{1/2}$ data are from:  \cite{Tiator:2003uu} (unshaded circles), \cite{Aznauryan:2004jd} (unshaded pentagons), \cite{Aznauryan:2005oral} (full circles), \cite{Armstrong:1998wg} (unshaded diamond); for $S_{1/2}$: \cite{Tiator:2003uu} (unshaded up triangles), \cite{Aznauryan:2004jd} (unshaded down triangles), \cite{Aznauryan:2005oral} (full triangles) }
\label{fig-S1535-AS}
\end{figure}

We obtain the form factors 

\begin{equation}
\begin{array}{l}
\di 
S_{11}(1535): \nonumber
\\
\di
g_1^{(p)}=\frac{2.0/D_V}{1+Q^2/1.2 M_V^2} 
	   \left[1+7.2\ln\left(1+ \frac{Q^2}{1\GeV^2}\right)
	   \right]\ ,		\nonumber
\\[4mm]
 \di
g_2^{(p)}=\frac{0.84}{D_V}
	   \left[1 + 0.11 \ln\left(1+\frac{Q^2}{1\GeV^2}\right)
	   \right]\ .
\label{ff-S1535}
\end{array}
\end{equation}

Notice here, that $g_2^p$ falls down slower than dipole (at least for $Q^2<3.5\GeV^2$),  supplying the most prominent contribution among the three isospin-1/2 resonances discussed in this paper. This  means experimentally, that the relative role of the second resonance region increases with  increasing of $Q^2$. Values of $Q^2=1-2\GeV^2$ are accessible (and are not suppressed kinematically) for $E_\nu\ge 1.5- 2 \GeV$. For these energies the $\SS$ and $\DD$ resonances are observable in the differential cross section.

The axial form factors  are determined from PCAC  as described in the  Appendix~\ref{gold-trei}:
\[
g_3^A{}^{(S)}=\frac{(M_R-m_N)m_N}{Q^2+m_\pi^2} g_1^A{}^{(S)},  \qquad g_1^A{}^{(S)}(0)=-0.21 .
\]
The $Q^2$ dependence is again taken 
\beq
g_1^A{}^{(S)}= \frac{g_1^A{}^{(S)}(0)/D_A}{1+Q^2/3M_A^2} .
\eeq
We adopt this functional form, but one must keep open the possibility that it may change when experimental results become available.

\section{Cross section for the tau neutrinos  \label{tau}}

Before we describe numerical results for the cross sections in the second resonance region, we shortly discuss the cross section for $\tau$--neutrinos and the accuracy achieved in different calculations.

Recently we calculated the cross section of the resonance production \cite{Lalakulich:2005cs} by taking into account the effects from the nonzero mass of the outgoing leptons. They generally decrease the cross section at small $Q^2$. For the muon the effect is noticeable in the $Q^2-$dependence of the differential cross section, but is very small in the integrated cross section.  

\begin{figure}[ht]
\begin{center}
\includegraphics[angle=-90,width=\columnwidth]{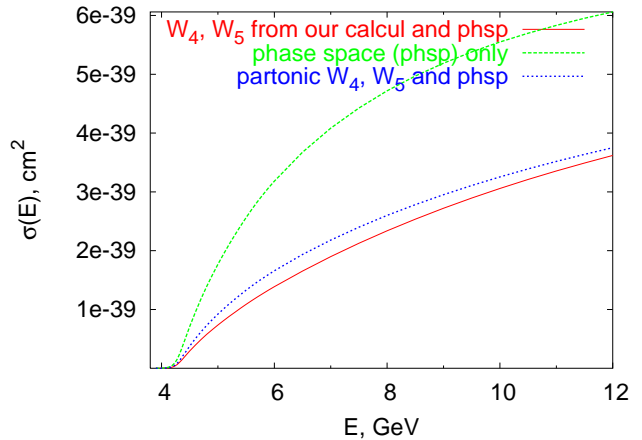} 
\end{center}
\caption{The integrated cross section $\sigma(E)$ as function of the neutrino energy. The dashed curve includes phase space corrections only; the dotted curve includes  phase space corrections and structure functions from the parton model; the solid curve includes phase space and structure functions calculated in our model.}
\label{fig-P1232-tau-integr}
\end{figure}

It was shown, that the cross section is decreased at small $Q^2$ via 1) reduction of the available phase space; 2) nonzero contributions of the ${\cal W}_4$ and ${\cal W}_5$ structure functions. Following \cite{Kuzmin:2004ya}, we'll refer to the latter effect as "dynamic correction". To date, several Monte Carlo Neutrino Simulators use the Rein-Sehgal model \cite{Rein:1980wg} of the resonance production as an input. In this model the lepton mass is not included. Thus, in Monte Carlo simulations the phase space is restricted simply by kinematics, but they do not take into account effects from ${\cal W}_4$ and ${\cal W}_5$ structure function. Some calculations are  also available, where the partonic values for the structure functions 
\beq
{\cal W}_4=0 \quad (a), \quad 
{\cal W}_5={\cal W}_2\cdot (\pq) / Q^2 \quad(b)
\label{calW-DIS}
\eeq
are included. We compare these two approximations with our full calculations for the integrated  and differential cross sections. Fig.~\ref{fig-P1232-tau-integr} shows the integrated cross section for the reaction $\nu_\tau p \to \mu^- \Delta^{++}$.

\begin{figure}[ht]
\begin{center}
\includegraphics[angle=-90,width=\columnwidth]{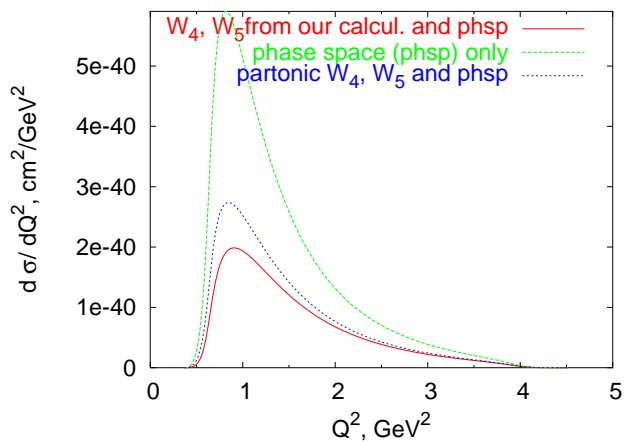} 
\end{center}
\caption{The differential cross section $d\si/d Q^2$ for the reaction $\nu_\tau p \to \mu^- \Delta^{++}$ at $E_\nu=5\GeV$. The meaning of the curves is the same as in Fig.~\ref{fig-P1232-tau-integr}.}
\label{fig-P1232-tau-Q2}
\end{figure}

One can easily see that taking the partonic limit for the structure functions is a good approximation in this case. Ignoring the structure functions, however, leads to a $100\%$ overestimate of the cross section which is inaccurate. In both cases the difference comes mainly from the "low" (close to the threshold) $Q^2$ region, as it is 
illustrated in  Fig.~\ref{fig-P1232-tau-Q2},  where the differential cross section for the tau neutrino energy $E_\nu= 5\GeV$ is shown. One easily sees, that the discrepancy at low $Q^2$ reaches $30\%$ for the partonic structure functions and more than $100\%$ when ${\cal W}_4$ and ${\cal W}_5$ are ignored.

\section{Cross sections in the second resonance region \label{xsec-2nd}}

\begin{figure}[ht]
\includegraphics[angle=-90,width=\columnwidth]{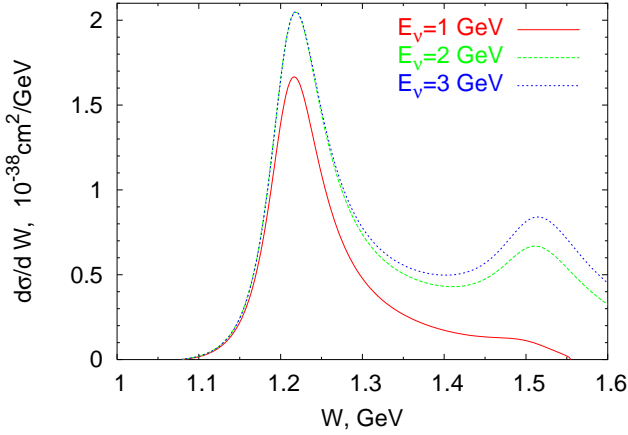}
\caption{Differential cross section $d\si / dW$ for the one-pion neutrinoproduction on neutron}
\label{dsidW-ppds}
\end{figure}

We finally return to the second resonance region and use the isovector form factors to calculate the cross section for neutrinoproduction of the resonances. We specialize to the final channels $\nu n \to R \to p \pi^0$ and  $\nu n \to R \to n \pi^+$, where both $I=3/2$ and $I=1/2$ resonances contribute. The data that we use is from the ANL \cite{Barish:1978pj,Radecky:1981fn}, SKAT \cite{Grabosch:1988gw} and BNL \cite{Kitagaki:1986ct} experiments. The ANL and BNL experiments were carried on Hydrogen and Deuterium targets, while the SKAT experiment used freon ($CF_3Br$). The experiments use different neutrino spectra, there is, however, an overlap region for $E_\nu<2.0 \GeV$. The data show that the BNL points are consistently higher that those of the other two experiments (see figure 4a,b in ref. \cite{Grabosch:1988gw}). This is also evident in earlier compilations of the data. For instance, Sakuda
\cite{Sakuda:2002rx} used the BNL data and his cross sections are larger that those of Paschos et al.\cite{Paschos:2002mb} where ANL and SKAT data were used. A recent article \cite{Sato:2003rq} uses data from a single experiment \cite{Barish:1978pj}, where the differences between the experimental results is not evident. The error bars in these early experiments are rather large and it should be the task of the next experiments to improve them and settle the issue.

The differential cross section $d \si / d W$ was reported in several experiments (see figures 4 in \cite{Barish:1978pj}, 1 in \cite{Radecky:1981fn}, 4 in \cite{Kitagaki:1986ct}, 7 in \cite{Allasia:1990uy}). We plot the differential cross section $d\si / d W$ in figure~\ref{dsidW-ppds} for incoming neutrino energies $E_\nu=1,2$ and $3\GeV$. We note, that the second resonance peak grows faster than the first one with neutrino energy and becomes more pronounced at $E_\nu=3 \GeV$. 

\begin{figure}[ht]
\includegraphics[angle=-90,width=\columnwidth]{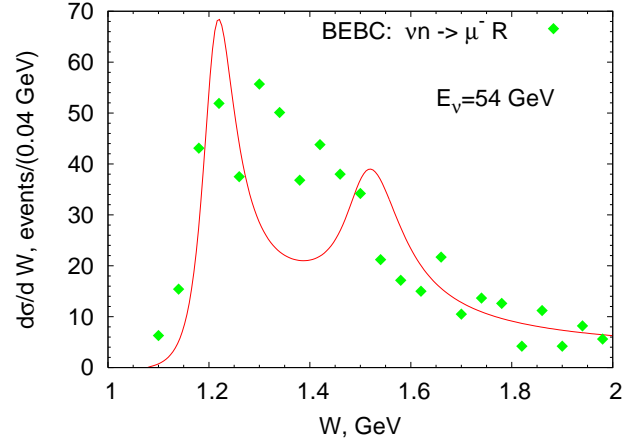}
\caption{Differential cross section $d\si / dW$ for the one-pion neutrinoproduction  for BEBC experiment}
\label{fig-BEBC}
\end{figure}

In Fig.\ref{fig-BEBC} we plot our theoretical curves together with the experimental data from the BEBC experiment \cite{Allasia:1990uy} for $ \langle  E_\nu \rangle =54\GeV$. The theoretical curve clearly shows two peaks with comparable areas under the peaks.  The experimental points are of the same order of magnitude and follow general trends of our curves, but are not accurate enough to resolve two resonant peaks.

The spectra of the invariant mass are also plotted in figure 4 in Ref.\cite{Kitagaki:1986ct} up to $W=2.0 \GeV$, but there is no evident peak at $1.4<W<2.0 \GeV$, in spite of the fact that the number of events is large. This result together with the fact that the integrated cross sections for $p\pi^0$ and $n\pi^+$ are within errors comparable suggest that the $I=1/2$ and $I=3/2$ amplitudes are comparable.

\begin{figure}[ht]
\includegraphics[angle=-90,width=\columnwidth]{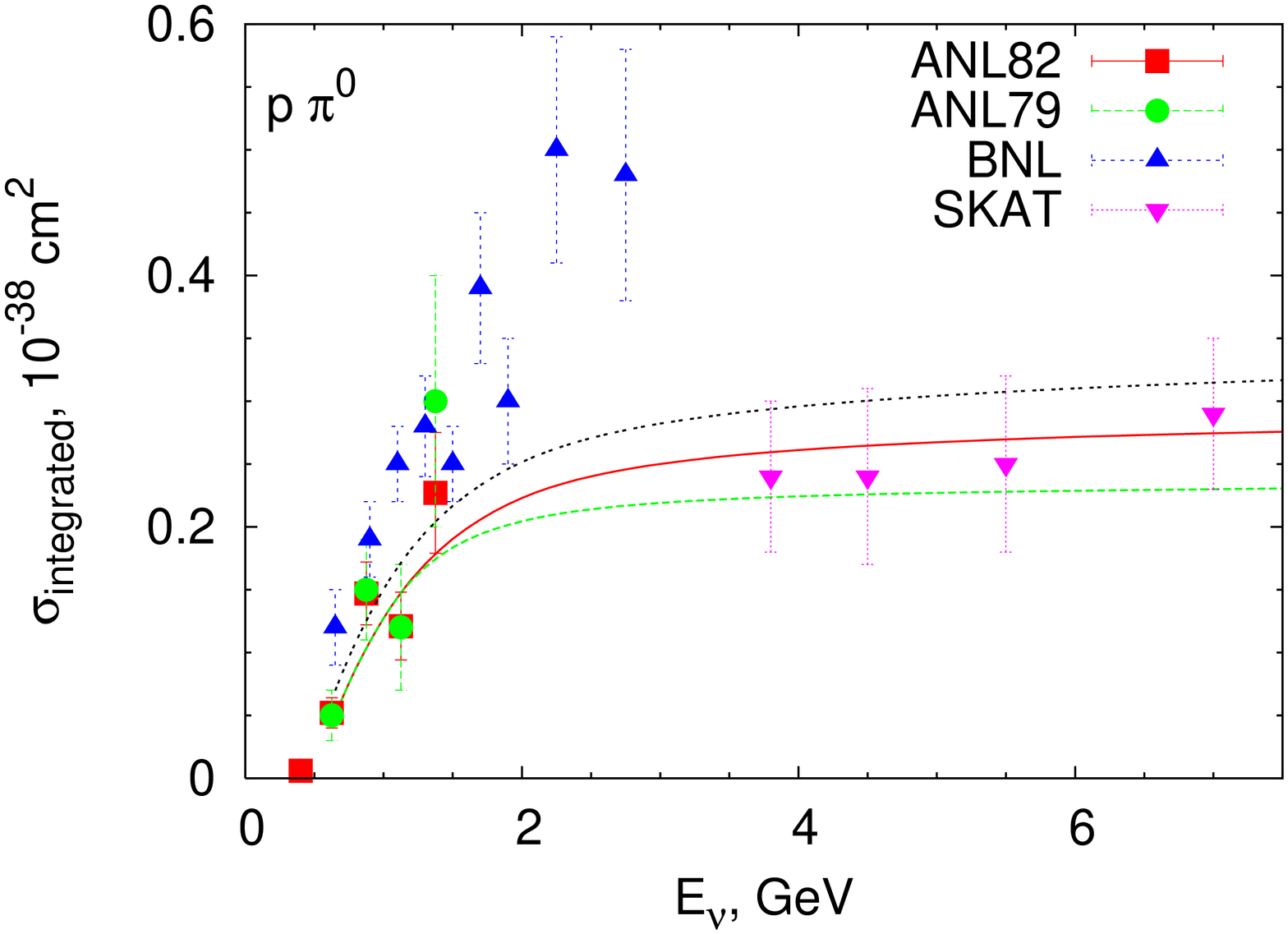}
\includegraphics[angle=-90,width=\columnwidth]{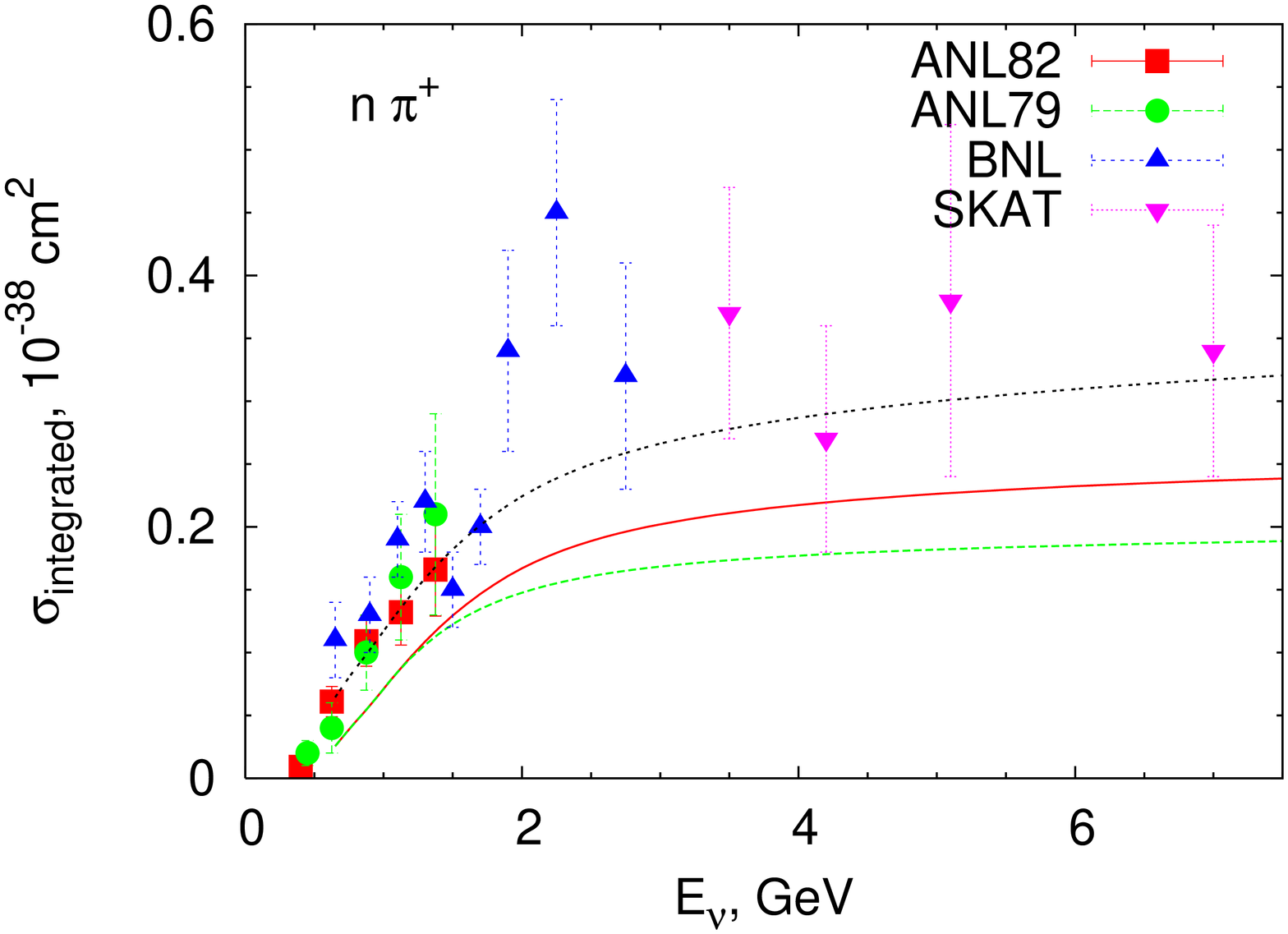}
\caption{Integrated cross section for the $\mu^- p \pi^0$ and $\mu^- n\pi^+$ final states}
\label{fig-si-integr}
\end{figure}

We study next the integrated cross sections for the final states $\mu^- p \pi^0$ and $\mu^- n\pi^+$ as functions of the neutrino energy. The solid curves in Fig.~\ref{fig-si-integr} show the theoretically calculated cross sections with the cut $W<2.0 \GeV$ and the dashed curve with the cut $W<1.6\GeV$. For $p \pi^0$ the solid curve goes through most of the experimental points except for those of the BNL experiment, which are consistently higher than those of the other experiments. 

For the $n\pi^+$ channel our curve is a little lower than the experimental points. This means that there are contributions from higher resonances or additional axial form factors. Another possibility is to add a smooth background which grows with energy. An incoherent isospin-1/2 background of approximately $5\cdot 10^{-40}(E_\nu/1\GeV-0.28)^{1/4}  \cm^2$ would be sufficient to fit the data, as it is shown by a double--dashed curve. By isospin conservation, the background for the $p \pi^0$ channel is determined to be half as big. Including this background, which may originate from various sources, produces the double-dashed curves in Fig.~\ref{fig-si-integr}. Since experimental points are not consistent with each other, it is premature for us to speculate on the additional terms. 

\begin{figure}[ht]
\includegraphics[angle=-90,width=\columnwidth]{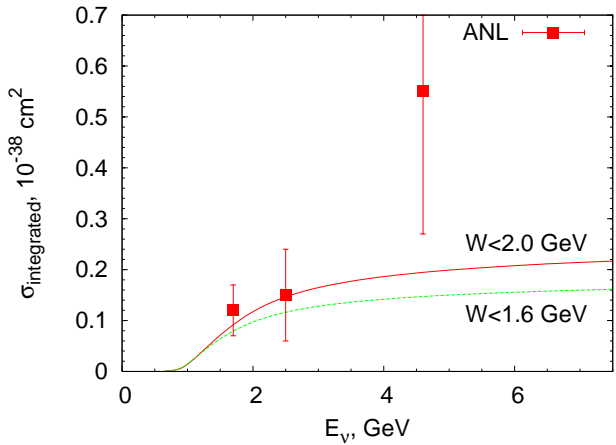}
\caption{Integrated cross section for the multipion production by neutrinos}
\label{fig-pds-multipion}
\end{figure}

For the $\PPP$ the elasticity is high, but for the other resonances is $\approx 0.5$, which implies substantial decays to multipion final states. We computed in our formalism the integrated cross section for multipion production. The results are shown in Fig.~\ref{fig-pds-multipion} with two cuts $W<1.6 \GeV$ and $W<2 \GeV$. The experimental points are from Ref.~\cite{Barish:1978pj}.

\section{Conclusions}

We described in this article a general formalism for analysing the excitation of resonances by neutrinos. We adopt a notation for the cross section very similar to that of DIS by introducing structure functions. We give explicit formulas for the structure functions in terms of form factors. The form factors describe the structure of the transition amplitudes from nucleons to resonances. The vector components appear in electro- and neutrino--production. We use recent data on helicity amplitudes from JLAB and the Mainz accelerator to determine the form factors including $Q^2$ dependences. We found out, that several of them fall slower than the dipole form factor, at least for $Q^2<2 - 3 \GeV^2$. The accuracy of these results is illustrated in Figures \ref{fig-D1520-AS}, \ref{fig-P1232-AS}, \ref{fig-P1440-AS}, \ref{fig-S1535-AS}. 

We obtain values for two axial form factors by applying PCAC (see Appendix~\ref{gold-trei}) whenever the decay width and elasticity is known. For the spin-3/2 resonances there is still freedom for two additional axial form factors whose contribution may be important.  This should be tested in the experiments.

We present differential and integrated cross sections in Sections \ref{tau} and \ref{xsec-2nd}. For the $\PPP$ we point out, that the structure functions ${\cal W}_4$ and ${\cal W}_5$ are important for experiments with $\tau-$leptons because they modify the $Q^2$ dependence and influence the integrated cross section. 

The second resonance region has a noticeable peak in $d\si / dW$ (Fig.\ref{dsidW-ppds}), which grows as $E_\nu$ increases from $1$ to $3\GeV$. The integrated cross section for the $I=1/2$ channel also grows with energy of the beam and may require stronger contribution from the resonance region and a non-resonant background (Fig.\ref{fig-si-integr}).

Our results are important for the new oscillation experiments. In addition to the production of the resonances and the decays to one pion and a nucleon, there are also decays to two and more pions. 
Multipion decays contribute to the integrated cross section with the cut $W<2\GeV$ at the level of $(2 - 3) \cdot 10^{-39} \cm^2$ for $E_\nu> 4 \GeV$.
Thus our results are useful in understanding the second resonance region and may point the way how the resonances sum up to merge at higher $Q^2$ into the DIS region.

\begin{acknowledgements}
We thank Dr. I. Aznauryan for providing us recent electroproduction data. The financial support of BMBF, Bonn under contract 05HT 4 PEA/9 is gratefully acknowledged. GP wishes to thank the 
Graduiertenkolleg 841 of DFG for financial support. One of us (EAP) wishes to thank the theory group
of JLAB for its hospitality where part of this work was done.
\end{acknowledgements}



\appendix

\section{Decays of the resonances and PCAC  \label{gold-trei}} 

\subsection{$P_{33}(1232)$}

For $P_{33}(1232)$ isospin invariance defines the following effective Lagrangian for the 
$\Delta N \pi$ interactions:
\begin{eqnarray}
{\cal L}_{\pi N R}^{P_{33}(1232)}= g_{\Delta}(\overline{\Delta^{++}_{\mu}} p \partial_\mu \pi^+ 
+\sqrt{\frac13} \overline{\Delta^{+}_{\mu}} n \partial_\mu \pi^+  
\nonumber \\
+\sqrt{\frac13} \overline{\Delta^{0}_{\mu}} p \partial_\mu \pi^-
+\sqrt{\frac23} \overline{\Delta^{+}_{\mu}} p \partial_\mu \pi^0  
\nonumber \\
+\sqrt{\frac23} \overline{\Delta^{0}_{\mu}} n \partial_\mu \pi^0
+  \overline{\Delta^{-}_{\mu}} n \partial_\mu \pi^-  )
\nonumber
\end{eqnarray}
The total width of the $\pi N$ decay of each $\Delta^{++}$, $\Delta^+$, $\Delta^0$ or $\Delta^-$ is calculated in a straight--forward way
\beq
\Gamma^{\Delta}=\frac{g_\Delta^2}{8\pi}\frac{1}{3 M_R^2} 
\left[ (M_R+m_N)^2-m_\pi^2 \right] |p_{\pi}|^3,
\label{Gamma-P33}
\eeq
where the pion momentum for the on--mass-shell resonance is
\[ 
p_{\pi}=\frac1{2M_R}\sqrt{(M_R^2-m_N^2-m_\pi^2)^2-4m_N^2 m_\pi^2} .
\]
For the experimental value $\Gamma_{\Delta}=0.114\GeV$, we obtain $g_\Delta=15.3 \GeV^{-1}$.
The resonance width (\ref{Gamma-P33}) is proportional to the third power of the pion momentum, so for the running resonance width we use 
\[
\Gamma^{(\Delta)}(W)=\Gamma_0^{(\Delta)}\left( \frac{p_\pi(W)}{p_\pi(M_R)} \right)^3,
\]

According to PCAC 
\beq
\langle R^+ | \partial_\mu A^\mu (0) |  n \rangle 
= -i m_\pi^2 f_\pi \frac{1}{q^2-m_\pi^2} T(\pi^+ n  \to R^+), 
\label{PCAC}
\eeq
where $f_\pi=0.97 m_\pi$.

For $P_{33}(1232)$ the relation (\ref{PCAC}) turns into
\[
i \overline{\psi_\mu^{\Delta+}} q^\mu \left[ C_5^A+\frac{C_6^A}{m_N^2}q^2 \right] u_N,
= \sqrt{\frac13}  \frac{-i m_\pi^2 f_\pi }{q^2-m_\pi^2}\overline{\psi_\mu^{\Delta+}} g_{\Delta} q^\mu u_N.
\]
and we obtain in the limit $m_\pi \to 0$ a relation between the two form factors 
\beq
C_6^A=-m_N^2 \frac{C_5^A}{q^2}.
\label{c5c6}
\eeq
The denominator of the above formula is phenomenologically extended  as $q^2-m_\pi^2$.
Making use of the relation (\ref{c5c6}) for $Q^2\to 0$ one also obtains $C_5^A=g_\Delta f_\pi /\sqrt{3}$.
Thus
\beq
C_6^A(P_{33})=m_N^2 \frac{C_5^A(P_{33})}{m_\pi^2+Q^2}, 
\quad 
C_5^A(P_{33})=\frac{g_\Delta f_\pi}{\sqrt{3}}=1.2
\label{P1232-C5A}
\eeq
For the $\Delta^{++}$ the $\pi N R$ vertex  is $\sqrt{3}$ times bigger, so, strictly speaking, $C_5^A$ is also $\sqrt{3}$ times bigger. However, for historical reasons, the form factors are conventionally defined for the vertex $W^+ n\to R^+$ and a factor $\sqrt{3}$ appears in vertex  for $W^+ n\to R^{++}$.


\subsection{$D_{13}(1520)$}

For $\DD$ the isospin-invariant Lagrangian of 
the $\DD N \pi$ interactions is defined as:
\begin{eqnarray}
{\cal L}_{\pi N R}^{D_{13}}= -g_{D13}\biggl[
 {\sqrt{\frac23}}  \overline{D^{+}_{\mu}} \gamma_5 n \partial_\mu \pi^+  
-{\sqrt{\frac23}} \overline{D^{0}_{\mu}} \gamma_5 p \partial_\mu \pi^-
\nonumber \\
- {\sqrt{\frac13}}\overline{D^{+}_{\mu}} \gamma_5 p \partial_\mu \pi^0  
+ {\sqrt{\frac13}} \overline{D^{0}_{\mu}} \gamma_5 n \partial_\mu \pi^0 .
\biggr]
\nonumber
\end{eqnarray}
The decay width to the $\pi N$  is 
\beq
\Gamma_{D \pi N}=\frac{g_{D}^2}{8\pi}\frac{1}{3 M_R^2} 
\left[ (M_R-m_N)^2-m_\pi^2 \right] |p_{\pi}|^3 .
\label{Gamma-D13}
\eeq
The total width of the $\DD$ resonances is approximately $0.125 \GeV$ and the elasticity  is about $0.5$.  For this values we obtain $g_{D}=15.5 \GeV^{-1}$ and the running width of the resonance is again proportional to the third power of the pion momentum.
The PCAC relation turns into
\[
i \overline{\psi_\mu^D} q^\mu \left[ C_5^A+\frac{C_6^A}{m_N^2}q^2 \right] \gamma_5 u_N
=  -\sqrt{\frac23} 
\frac{-i m_\pi^2 f_\pi }{q^2-m_\pi^2}
\overline{\psi_\mu^D} g_{D} q^\mu \gamma_5 u_N.
\]
which results in
\begin{eqnarray}
C_6^A(D_{13})&=&m_N^2 \frac{C_5^A(D_{13})}{m_\pi^2+Q^2},  
\label{D1520-C5A}
\\[2mm]  \nonumber
\mbox{with} \quad C_5^A(D_{13})(Q^2=0)&=&-\sqrt{\frac23}g_{D} f_{\pi} = -2.1 .
\end{eqnarray}

\subsection{$P_{11}(1440)$}

For $\PP$ the isospin-invariant Lagrangian is defined as
\begin{eqnarray}
{\cal L}_{\pi N R}^{P_{11}}= -g_{P11}\biggl[
  \sqrt{\frac23}\overline{P^{+}} \gamma_5 n  \pi^+  
- \sqrt{\frac23}\overline{P^{0}} \gamma_5 p  \pi^-
\nonumber \\
- \sqrt{\frac13} \overline{P^{+}} \gamma_5 p  \pi^0  
+ \sqrt{\frac13} \overline{P^{0}} \gamma_5 n \pi^0 .
\biggr]
\nonumber
\end{eqnarray}
The decay width is 
\[
\Gamma_{P11 \to \pi N} = \frac{g_{P}^2}{8 \pi M_R^2} 
\left[ (M_R-m_N)^2-m_\pi^2 \right] |p_{\pi}| .
\]
For the elasticity of $0.6$ and the total width of $0.350\GeV$ we obtain
$g_{P}=10.9$. 

The PCAC relation is
\[
\begin{array}{l} \di
i\bar{u}_R(p')  \left[ g_1^A \gamma^\mu q_\mu  + \frac{g_3^A}{m_N} q^2 \right] \gamma^5 u_N(p)=
\\[3mm] \hspace*{20mm} \di
=-\sqrt{\frac23} (-i m_\pi^2 )\frac{f_\pi}{q^2-m_\pi^2}\bar{u}_R(p') g_{P} \gamma^5 u_N(p) . 
\end{array}
\]
At  $m_\pi \to 0 $ it leads to 
\[
g_3^A(P_{11})=-\frac{m_N(M_R+m_N)}{q^2-m_\pi^2} g_1^A(P_{11})
\]
(here the denominator is extended as before)
and at $Q^2\to 0$ the coupling is
\[ \di
g_1^A(P_{11})=-\sqrt{\frac23} \frac{g_{P} f_\pi}{M_R+m_N}=-0.51
\]

\subsection{$S_{11}(1535)$}

For $\SS$ the isospin-invariant Lagrangian is defined as
\begin{eqnarray}
{\cal L}_{\pi N R}^{S_{11}}= -g_{S}\biggl[
 \sqrt{\frac23} \overline{S^{+}}  n  \pi^+  
-\sqrt{\frac23} \overline{S^{0}}  p  \pi^-
\nonumber \\
- \sqrt{\frac13} \overline{S^{+}}  p  \pi^0  
+ \sqrt{\frac13} \overline{S^{0}}  n  \pi^0 .
\biggr]
\nonumber
\end{eqnarray}
The decay width  is 
\[
\Gamma_{S11 \to \pi N} = \frac{g_{S}^2}{8 \pi M_R^2} 
\left[ (M_R+m_N)^2-m_\pi^2 \right] |p_{\pi}| .
\]
For the elasticity of $0.4$ and the total width of  $0.150\GeV$ we obtain
$g_{S}=1.12$.

The PCAC relation is again
\[
\begin{array}{l} \di
i\bar{u_R}(p') \left[  g_1^A \gamma^\mu q_\mu  + \frac{g_3^A}{m_N} q^2 \right] u_N(p)=
\\ \hspace*{20mm} \di
=-\sqrt{\frac23}(-i m_\pi^2) \frac{f_\pi}{q^2-m_\pi^2}\bar{u}_R(p') g_{S} u_N(p)
\end{array}
\]
which at $m_\pi \to 0 $ leads to 
\[
g_3^A(S_{11})=-\frac{m_N(M_R-m_N)}{q^2-m_\pi^2} g_1^A(S_{11}) .
\]
(here the denominator is extended as before)
and at $Q^2\to 0$ the coupling is 
\[ \di
g_1^A(S_{11})=-\sqrt{\frac23}\frac{ g_{S} f_\pi}{M_R-m_N} = -0.21 .
\]

\bibliographystyle{apsrev}
\bibliography{references}

\end{document}